\documentclass[sigplan,10pt]{acmart}
\renewcommand\footnotetextcopyrightpermission[1]{}

\usepackage{graphicx}
\usepackage{balance}  
\usepackage{cleveref}
\usepackage[inline]{enumitem}
\usepackage{url}
\usepackage[utf8]{inputenc}
\usepackage{booktabs}
\usepackage{subcaption}
\usepackage{xcolor}
\usepackage{hyperref}
\hypersetup{
    colorlinks=true,
    linkcolor=black,
    filecolor=black,      
    urlcolor=black,
    citecolor=black,
}
\usepackage{float}

\AtBeginDocument{%
  \providecommand\BibTeX{{%
    \normalfont B\kern-0.5em{\scshape i\kern-0.25em b}\kern-0.8em\TeX}}}

\begin{document}
\pagestyle{plain}

\title{VLDB 2021: Designing a Hybrid Conference}

\author{Pınar Tözün}
\affiliation{%
  \country{IT University of Copenhagen}
}
\email{pito@itu.dk}

\author{Felix Naumann}
\affiliation{%
  \country{Hasso Plattner Institute, University of Potsdam}
}
\email{felix.naumann@hpi.de}

\author{Philippe Bonnet}
\affiliation{%
  \country{IT University of Copenhagen}
}
\email{phbo@itu.dk}

\author{Xin Luna Dong}
\affiliation{%
  \country{Facebook}
}
\email{lunadong@fb.com}

\begin{abstract}
  In 2020,
  while main database conferences one by one \cite{BonifatiGLMMPSS20, MaierPDDHLLMRST20}
  had to adopt a virtual format as a result of the ongoing COVID-19 pandemic,
  we decided to hold VLDB 2021 in \textit{hybrid} format.
  This paper describes how we defined the hybrid format for VLDB 2021
  going through the key design decisions.
  In addition, we list the lessons learned from running such a conference.
  Our goal is to share this knowledge with fellow conference organizers
  who target a hybrid conference format as well,
  which is on its way to becoming the norm rather than the exception.
  
  For readers who are more interested in the highlights rather than details,
  a short version of this report appears in SIGMOD Record \cite{BonnetDNP21}.
\end{abstract}

\settopmatter{printacmref=false}
\maketitle

\section{Defining the Hybrid Format}
\label{sec:hybrid}

The 47th International Conference on Very Large Databases (VLDB) 2021 was held as a hybrid conference in Copenhagen, Denmark on August 16-20, 2021.
It attracted $\approx$180 in-person attendees in Copenhagen and $\approx$840 remote attendees.
After several virtual conferences during the pandemic,
this was the first major conference that brought together the database community in person.
It was also the first attempt at organizing a hybrid conference at this scale in the community.

We faced several key questions to answer to settle on
what the hybrid format means for VLDB 2021.
The following subsections discuss the design space and our choices. 

\subsection{Conference Schedule}
\label{sec:hybrid:conference}

An initial question was whether
to adopt a 24-hour cycle while having some local in-person component in Copenhagen or
to primarily focus on the Copenhagen time while providing some global accessibility.

VLDB 2020, which was held as a virtual-only conference,
implemented the 24-hour format by splitting the conference into 4-hour blocks
and repeating sessions to allow attendees from all timezones to catch
all the sessions at a reasonable waking time.
SIGMOD 2021, in contrast, adopted the 24-hour format with repeated sessions but
using less uniform and longer time-blocks.
Also, SIGMOD 2021 had a local in-person component in Beijing
while designing a primarily virtual conference,
since most of the attendees had to be virtual. 

Based on the experiences from VLDB 2020, SIGMOD 2021 and other recent conferences,
the question of scheduling boiled down to
whether to place the virtual component first or the in-person one.
Our goal for VLDB 2021 was to have a strong in-person component from the beginning.
Therefore, we decided to focus on the Copenhagen timezone by
having all components of the main conference program
at a suitable local time. 
Nevertheless, the hybrid format requires creating a good experience for the virtual participants as well.
Therefore, we deviated from the traditional conference schedule structure in the following ways:
\begin{list}{\labelitemi}{\leftmargin=1.5em}
\item{Keynotes were scheduled in the afternoons in Copenhagen to be
more accommodating for international timezones, both in Asia and in the Americas.}
\item{Poster and demo sessions were each scheduled once in the early morning and once in late evening timeslots in Copenhagen time in a virtual-only setting, to allow the authors from different timezones to attend at their convenience.
This decision simplified the costs and complexity of (1)~hybrid poster and demo sessions
and (2)~ensuring a safe setup for such sessions during the pandemic.}
\item{We included virtual-only roundtable sessions
because of their popularity and effectiveness in previous virtual-only conferences.}
\end{list}

These key decisions for the main conference schedule structure
impacted the rest of the conference design.

\subsection{Workshop schedule}
\label{sec:hybrid:workshops}

The key question for scheduling workshops is their level of flexibility.
Do we allow some workshops to have only a virtual component
or enforce the hybrid format for all workshops?
Do we allow the workshop chairs to define the timezone they wish to optimize for,
or do we set Copenhagen-based time boundaries for them?

With our priority of the in-person component and Copenhagen time for the main conference,
we followed the same high-level principle for workshops,
while giving the workshops certain flexibility:
First, while we encouraged workshops to run in hybrid mode,
we gave the option of organizing virtual-only workshops
if the workshop chairs preferred this option.
This was necessary at the time,
considering the travel restrictions and health concerns due to the ongoing pandemic.
In the future, even without an ongoing pandemic,
we may want to give such flexibility to workshops
to allow flexible attendance.
In the end, 6 out of the 13 workshops were virtual-only, while the rest ran in hybrid mode.

Regardless of the virtual-only or hybrid setup,
the workshops should integrate well with the main conference.
Therefore,
we determined time boundaries for the workshop schedules, to assist in-person attendees.
For the virtual-only workshops, this was from 8am to 10pm Copenhagen time.
For the hybrid ones, this was from 8am to 7pm Copenhagen time, based on room availability.

Some hybrid workshops ended up starting early in the day and finished early afternoon,
while others shifted their start time to 10am and ended slightly later.
Workshops with virtual components had a wider variety in their start and end times.
Some adopted times similar to hybrid workshops,
while others chose to run during Copenhagen late afternoon and evening time.
Overall, each workshop chair had the chance to accommodate as much as possible the timezones of their attendees, while still being associated with the preferred timezone of the conference.

\subsection{Technical Platforms}

Online platforms are crucial for conferences that have a virtual component.
Therefore, an important design choice was the set of platforms to support the conference.

Most recent virtual conferences chose a
combination of platforms (Zoom, Slack, Gather, etc.) instead of relying on just one.
On the one hand,
using too many platforms can become cumbersome for both the organizers and attendees.
On the other hand, the platforms should
cover all needs of a hybrid conference,
such as keeping an easy-to-follow schedule across timezones,
supporting the availability of session presentations virtually,
enabling discussions among all attendees,
boosting interactions across in-person and virtual parts of the conference,
and giving sponsors good options to reach the attendees.
In addition, one should keep in mind both the quality of the platforms and
the familiarity of the conference attendees with a particular platform.

Based on the considerations above,
our main design goal was to simplify the conference experience
(finding information about the sessions, interacting with other attendees, etc.)
for virtual attendees.
We did this by
(1)~minimizing the total number of platforms to be used
by choosing the \textit{Whova} event management software
as the only entry point to the virtual part of the conference, and
(2)~choosing \textit{Zoom},
which is by now extremely familiar to almost everyone,
to support streaming the sessions to virtual attendees.
In addition, as usual across all conferences,
we also used the conference webpage to keep the conference schedule information publicly
and other important conference announcements,
such as student attendance awards, venue information, etc.
Finally, we also used YouTube and Bilibili for uploading the pre-recorded videos.

In Whova,
to boost interactions between the sponsors and the attendees
and the paper authors and the attendees,
we enabled the add-ons \textit{exhibitor center} and \textit{artifact center}, respectively.
The exhibitor center allows sponsors to customize their interactions with the attendees,
while the artifact center allows paper authors to continue discussions
beyond the sessions with other attendees.

In Zoom, we opted out of the webinar mode to further increase interactions among the attendees.
While the webinar mode of Zoom is more secure against disruptive attendees,
it creates an isolating experience for both the attendees and the presenters.
The conference organizers have the power to react to disruptions,
rather than being pessimistic and pro-actively avoiding them,
especially when conference access requires a paid registration.

\subsection{Sessions}

Based on the decisions made on the schedule, we supported two types of sessions:
\textit{hybrid} and \textit{virtual-only}.
The choice of virtual platforms also impacted their design.

\subsubsection{Hybrid sessions}

As hybrid sessions we scheduled all research and industry paper sessions, tutorials, sponsor talk sessions, and the (semi-)plenary sessions with keynotes, panel, award session, and scalable data science talks.
The key design question was whether to hold these presentations live or pre-recorded.
To create a lively experience, we encouraged speakers to give live talks, regardless of their in-person or virtual attendance. Nevertheless, we collected pre-recorded 10-min videos for all papers as backup and for archival purposes, and gave speakers the option of using their pre-recorded talk.
All but one in-person speakers and the other attendees were glad to have the experience
of live in-person talks after a very long time.

The virtual speakers had three options:
(1)~giving a live talk over Zoom and being present for Q\&A afterward,
(2)~using their pre-recorded video over Zoom and being present for Q\&A afterward,
(3)~using their pre-recorded video, but being absent for Q\&A due to timezone mismatch.
To encourage our preferred option -- the first one -- we gave a slightly longer presentation time (15~mins instead of the 10~mins pre-recorded video) to live presenters.

In the end, across all research and industry sessions, $\approx$
20\%, 48\%, and 15\% of the speakers picked options 1, 2, and 3, respectively.
Rest of the talks were in-person live in these sessions.
In contrast, keynotes, award talks, and scalable data science talks
were all live either in-person or over Zoom.

Another key design question was whether to organize the paper sessions primarily
based on the topic of the papers or the timezone and attendance mode of the speakers.
We chose to group the papers based on topic first
in order to have a more natural flow in each session
and guide the attendees to the presentations they would be most interested in.
Then, we attempted to schedule each topic-based paper session at a timeslot
that is the most ideal for the majority of the speakers in that session,
based on their timezone.
While this worked well for many speakers, there were some cases,
where the timeslot was not ideal for the paper authors.
This was the main reason we created the third presentation mode option from above.
We aimed at compensating for such scenarios through the interactions
during poster sessions and at the artifact center in Whova.
Unfortunately,
such interactions ended up being low during the conference
as we will cover in more detail in Section~\ref{sec:lessons}.

The hybrid sessions also require additional coordination.
The information on the presentation modes for all the talks
were collected ahead of time shortly before the conference start date
by the conference organizers,
and were distributed to all the necessary parties such as
the audio-visual (AV) personnel streaming the in-person part of the conference over Zoom,
the Zoom technician managing the Zoom sessions,
and the session chairs coordinating both the in-person and virtual speakers and attendees.

\subsubsection{Virtual-only sessions}
\label{sec:hybrid:sessions:virtual}

The virtual-only sessions of the conference were the poster, demo, and roundtable sessions.
The key design question here was
how to integrate these sessions with the virtual platforms of the conference.

The poster and demo sessions were designed as parallel Zoom sessions,
with a breakout room dedicated to each paper poster and demo.
The navigation information for finding a particular paper and demo was on Whova.
There were also session chairs monitoring one or more of these Zoom sessions to ensure
everything is running smoothly and to encourage interactions among the attendees.

Each roundtable session had their dedicated Zoom session and one or more session chairs.
The topics for the roundtables were determined by the session chairs before the conference,
and the session chairs had full freedom in terms of
how they wish to organize and run these sessions.

\subsection{Social events}

The in-person social events are among the most valuable parts of a conference,
since they bring the community together.
Virtual-only conferences lack such in-person opportunities.
The key design question here was
whether and how to adapt the social events to the conference schedule of the hybrid format.
Should they target just the in-person attendees who traveled all the way to the conference?
Or is it possible to design social events that include both in-person and virtual attendees?

For the banquet, we followed the traditional format:
only for the in-person participants on Wednesday evening.
The evening online poster and demo sessions
were omitted for that evening to avoid a clash with the banquet.

For the opening reception, on the other hand, we followed a different route.
First, we moved it from its typical slot of Monday evening to Tuesday evening,
since there were some Monday workshops running during Copenhagen evening time
as a result of our flexible schedule for the workshops (Section~\ref{sec:hybrid:workshops}).
Then, we changed its focus to promote the D\&I initiative
across data management conferences \cite{dnidb}.
Finally, we coupled this event with a D\&I keynote following the focus of the event.
This combination of the reception and keynote turned this social event to a hybrid social
experience rather than just an in-person one.

\subsection{Registration}
\label{sec:hybrid:registration}

Designing the registration process for a hybrid conference requires more than just adding additional categories to the conference registration options and determining their corresponding costs.
In particular, we needed to allow for some flexibility:
attendees decided late whether to attend in person in light of the fluctuating pandemic situation.

We first decided that there will be no free registrations even for the virtual attendees except for the sponsors, keynote speakers, and student attendance grant receivers.
While this is less inclusive for virtual attendees, the hybrid format brings additional costs, such as the cost of streaming live in-person sessions for the virtual attendees.
These costs should not burden only the paper authors,
since they are among the ones contributing to the attractive content of a conference.
In addition, introducing a fee even for the virtual attendees was expected minimize the number of people who register for the conference, but do not attend in the end.
In addition and as usual, we designed the overall categories and prices in a way that keeps the cost for student attendees as low as possible to promote student participation.

Finally, we chose to have a flexible registration process that allows \emph{upgrading} from virtual registration to in-person one and \emph{downgrading} from in-person registration to virtual one up to two weeks before the conference start date.
In this way, attendees could accommodate for change in plans, especially under the many unknowns created by the still ongoing COVID-19 pandemic.
The downside of this flexibility is that it becomes difficult to account for how many people will be at the conference, even at dates very close to the conference.
Therefore, the cut-off dates for such flexibility should be determined based on the deadlines associated with catering, social events, bus transfers, etc.

\subsection{Volunteers vs.\ professional help}

The question of whether to rely on volunteers or paid professionals to perform a particular task associated with the conference is relevant for all types conferences.
The number of tasks to manage, however, 
is naturally higher for a hybrid conference,
due to the increased number of platforms to keep track of
and people to inform when changes happen.

We received professional help for audio/video (AV), Zoom, and video collection and upload.
Getting professional help for the audio/visual personnel is a must.
While the management of Zoom sessions and pre-recorded videos can be done by volunteers,
we used \textit{Gateway}'s services, similar to SIGMOD 2020 and VLDB 2020.
Gateway knows our conferences and the virtual platforms such as Zoom, Whova, and YouTube very well at this point, and their services were extremely valuable during the conference.

On the other hand, we had student volunteers organized by our digital platform and artifact chairs to populate the content in Whova and upload videos to Bilibili, respectively.

Finally, we had both professional and volunteer help for the registration process.
We relied on the professional conference organizer \textit{Kuoni}
to manage the official registration page.
This was especially needed to handle the upgrades and downgrades mentioned
in Section~\ref{sec:hybrid:registration},
since Whova's built-in registration tool does not allow such flexibility. 
Therefore, our digital platform chairs had to 
periodically transfer the list of registered people
from Kuoni's database to Whova.


\section{Lessons Learned}
\label{sec:lessons}

This section covers our lessons learned running VLDB~2021 based on
the hybrid format defined in the previous section.

\subsubsection*{It is more important to be ready to recover quickly
    rather than to expect everything will run perfectly.
    It is possible to be agile during the conference.}
Even with the best plan and pre-conference testing,
there will be failures during a conference,
especially a big premier conference like VLDB
with typically more than 800 attendees and multiple parallel sessions.
A hybrid conference,
compared to more traditional in-person or virtual-only conferences,
have more things to keep track of, which amplify this situation.
Therefore,
we recommend the conference organizers to be ready for a swift recovery
and even be agile and change some key decisions during the conference as needed.

During the paper sessions unexpected incidents happened,
such as Zoom links not working,
speakers changing the presentation format,
a backup video not being available at the time it was supposed to be played,
etc. 
In all these scenarios,
the situation was fixed rapidly by the help of
our Zoom technicians (fixing the links and finding corresponding videos)
and session chairs (reordering content of the session).
The conference continued running smoothly despite such minor hiccups.


\subsubsection*{A simpler setup is better and less risky
    than a fancier and more complex one when it comes to AV.}
Our initial plan for the AV setup had a mechanism
that combined the camera image of the in-person speakers
with the image of their projected slides,
and then this combined image would be streamed over Zoom.
In this way, the in-person speakers did not have to worry about
connecting to Zoom and share their presentation over it.
We did test this setup the week before the conference with the AV team from the venue.

As the first sessions of the conference started with the Monday workshops, the video streamed based on the mechanism above turned blurry after a while.
The streamed video was too heavy for the network at the venue.
To solve this problem,
we quickly decided to switch to a mode
where each in-person speaker connected to Zoom sessions
and shared their presentation themselves.
The AV personnel now just had to share the speaker's video, which is lighter, 
instead of the combined stream of this video and the projected slides.
The in-person speakers did not have difficulty connecting to corresponding Zoom sessions,
since the Zoom links were easily available at Whova
and almost everyone had Zoom installed on their work laptops.

\subsubsection*{
Individual breakout rooms for each poster and demo impede interaction.}
As Section~\ref{sec:hybrid:sessions:virtual} describes,
we designed the poster and demo sessions
to be virtual-only and ran them as parallel Zoom sessions
with individual breakout rooms for each paper, poster, and demo.

At the Monday evening and Tuesday morning demo sessions,
we realized that this part of the conference had very few attendants
(less than five attendees excluding the authors)
and interactions were low.
During the Tuesday evening poster session, the situation was even more dire,
considering the higher number of parallel sessions and content.
Even the majority of the paper authors were absent from the session.
As a result, midway through this poster session,
we decided to turn these parallel Zoom rooms into roundtables, removing the breakout rooms for the papers and demos.
We deployed either one main room or two breakout rooms
depending on the number of participants in the Zoom session.
Each paper representative in the session
could introduce themselves and their work briefly,
and the others could ask questions to them.
We also dedicated a senior member to chair the discussion during the roundtable.
Overall, the poster session turned into a networking roundtable,
especially for the student authors,
where there was a lot more interaction compared to the initial setup.
We adopted this roundtable format for the Thursday poster session as well.

In future hybrid conferences,
organizing such networking roundtables could be helpful
for virtual student attendees to network with each other.
For virtual poster and demo sessions,
one can investigate more interactive options such as Gather 
or incorporate demos into the paper sessions as done in SIGMOD 2021. 
The most effective approach to virtual posters and demos is yet to be seen.


\subsubsection*{Low parallelism during roundtables is better.}
Round-table sessions have become a very popular part of VLDB and SIGMOD
recently, leading to many fruitful discussions.
For VLDB~2021,
we had many exciting roundtables lined up by our roundtable chair,
and chose to run them in parallel (4 to 7 at a time).
However, this parallelism hurt the attendance.
For future conferences, having not more than a couple of roundtable sessions in parallel should be a design principle to increase attendance and interactivity.

\subsubsection*{The role of session chairs is even more important during hybrid sessions.}
Session chairs play an important role in any conference.
A hybrid session setup increases the responsibilities of session chairs.
First, they have to give directions to speakers and attendees about the hybrid setup,
such as informing them about where to stand and speak into the microphone to be audible and visible.
Then, they have to monitor both in-person and virtual attendees to prevent people from being disruptive for the session.
Finally, they have to bridge the in-person and virtual parts of the session by coordinating speakers and questions on both sides.

In light of these increased set of responsibilities, it is worthwhile to aim at having two session chairs at each session -- one playing the traditional role of a session chair, the other acting as a stand-in for the online-participants, monitoring their questions. With the lower number of senior researchers attending in person, we were forced to ask them to chair two or even three sessions, and if possible, recruit a second session chair before the session began. 
For sessions with few attendees, one session chair proved to be sufficient.

\subsubsection*{In a hybrid conference,
    there is need for a digital platform chair, artifact chair, and scheduling chair.}
The additional complexity a hybrid conference brings requires additional conference roles.
While we established the first two of these roles from the start for VLDB~2021,
the need for the third one emerged
as we were entering the last stretch of our conference preparation.

The role of the \emph{digital platform chair}
has already become part of our conferences with the virtual format. 
This role is necessary to manage the content on the virtual event platform and
coordinate with other parties about population of and updates on this content.
Digital platform chairs are also the first responders
when the conference attendees have questions about
the virtual platforms of the conference.

The role of the \emph{artifact chair} is essential to manage
the process for collecting research artifacts, such as
pre-recorded videos, posters, etc.,
and coordinating the parties that are involved,
from paper authors to Gateway (in VLDB~2021).
After the conference,
the artifact chair hands off the necessary information and consents
to the person(s) responsible for archiving these artifacts.

The role of a \emph{scheduling chair} emerged
due to the need to manage a program
with a lot of metadata.
The schedule for a traditional conference program mainly contains
information about which paper is presented or who presents at each session.
This information is also enough for the attendees to decide which sessions to attend.
However, to be able to run the sessions of a hybrid conference,
the \emph{ground-truth schedule document} must be created
that contain the additional information of
the Zoom links, video location information for pre-recorded videos,
the presentation modes for each talk, etc.
This requires integration of information from several parties.
The general chairs can then decide who sees what part of this information
among the organizers, session chairs, and the attendees.

\subsubsection*{Finding an ideal set of virtual platforms
    for our future hybrid conferences may take some time.
    Improving interaction
    between virtual and in-person attendees is a priority.}
Our community is still trying out different virtual platforms for our conferences.
Everyone has their preferences.
Some like having access to everything through a single platform like Whova.
Some instead prefer having Slack or Gather as additional platforms
to increase interactions among conference attendees.

In VLDB~2021, 
the interactions among in-person attendees and
the interactions among the virtual attendees
during plenary or semi-plenary sessions were very fruitful.
However,
the setups we had over Whova to boost interactions across all attendees,
such as exhibitor and artifact centers,
were not as highly used as we envisioned.
In the future,
we recommend scheduling more networking sessions
and deploying an additional platform such as Slack
to boost interactions.

\subsubsection*{Session recordings should be available as soon as possible.
}
We decided to record all sessions,
unless requested otherwise by the speakers,
and make them available to the attendees through Whova.
However, with our program being longer than a daily program of
a traditional in-person conference with up to seven parallel sessions,
it took some time for Gateway to edit and upload these session recordings to YouTube.
As a result, most sessions recordings were made available from Wednesday on.

During the delay period,
we received many requests from the (virtual) attendees for the session recordings.
It is clear that the attendees who miss sessions of interest
because of either timezone conflicts or other forms of unavailability,
really value being able to catch up speedily with these sessions afterward.
Therefore,
we also decided to make these recordings available in Whova
for a while longer after the conference ended.

In the future, it would be crucial to either
arrange a setup where the session recordings are available
directly after the corresponding sessions are over,
or communicate clearly to the attendees when the recordings will be available.

\subsubsection*{Conferences should establish an archival
    procedure that is GDPR-compliant
    for the pre-recorded videos and session recordings.}
As mentioned above,
making session recordings available during and shortly after
the conference is very valuable for the conference attendees.
However, there is also high demand and value for long-term archiving (at least some of) these recordings.
How to archive them in a GDPR-compliant manner is,
on the other hand, a non-trivial challenge.

The session recordings tend to have several attendees appear
in the recording for brief moments of time in addition to the speakers.
Getting consent from everyone who appears in the recordings may not be feasible.
Unfortunately, for VLDB~2021,
we could not devise a scalable plan for the GDPR-compliant archival of session recordings.
The pre-recorded videos of papers will be archived, though, on the PVLDB website,
since they do not pose the same level of complexity for GDPR\@.
Simply receiving consent from the speakers is enough to archive them in a GDPR-compliant way.

There are several steps we can take to ease the challenge of
archiving session recordings in our future conferences.
We can collect the necessary consents for this during registration.
However, according to GDPR, 
anyone has the right to take back their consent at any point in time.
Therefore, we also must establish a process to handle such cases,
by either removing the corresponding videos
or editing out the corresponding person.
In order to make this process more scalable,
while recording the sessions, we should only capture the minimum necessary.
This would mean capturing the session chairs, presentations, speakers,
and (maybe) people who ask questions, for instance by using Zoom's speaker mode for the recording.
Minimizing the number of persons identifiable in the recording
would in turn minimize the number of consents to manage.
It may even obviate the need for collecting consents during registration.
The session chairs and speakers in a session are already known from the conference program,
and the session chairs or a volunteer can keep track of people who ask questions.

\subsubsection*{We should establish guidelines and initiatives for sustainability of our conferences.}
While the topic of sustainability may seem orthogonal to the hybrid conference design,
the hybrid format has great potential to create more sustainable conferences,
while making the conference accessible to a wider audience.
%
As investigated by our \textit{sustainability chair},
we can establish certain sustainability guidelines and initiatives
for future conference organizers.
A guideline could be to avoid food and plastic waste or to review the sustainability principles of the conference venue.
An initiative could be to offset the estimated carbon footprint of the conference
using any surplus of that year.

\subsubsection*{The hybrid setup harbors more unknowns and risks, even without a pandemic.}
One risk involves attendance numbers. For the traditional in-person conferences, 
we have a rough idea of how many people would attend.
On the other hand, for the hybrid format,
we have not yet established good estimations
on the in-person and virtual attendance numbers also based on conference location.
Even without a pandemic causing travel restrictions and health concerns,
these numbers can fluctuate, making planning more challenging for the conference organizers.
Flexible catering options,
establishing a price for virtual attendance, and
creating an overall schedule slightly earlier to plan for AV costs in detail
are among the things to keep in mind for reducing financial risks despite the unknowns.

\subsubsection*{The pandemic requires additional conference roles and responsibilities.}
While organizing a hybrid conference is orthogonal to the impact of the pandemic,
conference organizers should be prepared for this impact as it may not go away in the near future.
There must be dedicated conference roles for keeping track of
both the local and global rules and regulations regarding the COVID-19 pandemic, and post the necessary information on the conference website. 

In addition,
the organizers must decide whether to just follow the rules of the host country or to establish stricter rules for testing and vaccination.
In VLDB~2021, we decided to follow the rules and regulations of Denmark,
which were sufficient and worked well.
The only additional activity we organized was
a test convoy 
for the people who were not yet vaccinated and needed a negative test to enter the banquet. 

Finally,
the organizers must also establish what to do
when someone is tested positive for COVID-19 or
was in close contact with someone who tested positive.
We prepared guidelines following the ones of the Danish government,
and luckily, we only needed it for the latter scenario.

\section{Conclusions}

The hybrid format for conferences is here to stay and opens up new opportunities for everyone.
On the one hand,
they are more inclusive by allowing people who are unable or unwilling to travel
still be part of the conference.
On the other hand,
they still serve to bring the research community together in-person.
However, it is also a new format that we are all in the process of defining.
This paper summarized our experiences with
defining the hybrid format and running a hybrid conference for VLDB~2021.
We hope these experiences help fellow conference organizers
aiming at organizing a hybrid conference in the future.

\section*{Acknowledgements}

As VLDB 2021 organizers,
we are grateful for the support we received from our community throughout.
We would like to thank
the VLDB Endowment for the support,
all the conference officers and volunteers for their extremely hard work, and
all the attendees for their constructive collaboration and feedback.

\bibliographystyle{ACM-Reference-Format}
\bibliography{vldb2021}

\end{document}